\documentclass[12pt,preprint]{aastex}
\begin{document}
%\submitted{}
\title{Neutrino-Induced Fission and $r$-Process Nucleosynthesis}
\author{Y.-Z. Qian}
\affil{School of Physics and Astronomy, University of
Minnesota, Minneapolis, MN 55455; qian@physics.umn.edu}

\begin{abstract}
An $r$-process scenario with fission but no fission cycling is
considered to account for the observed abundance patterns of
neutron-capture elements in ultra-metal-poor stars. It is 
proposed that neutrino reactions play a crucial role in inducing 
the fission of the progenitor nuclei after the $r$-process freezes 
out in Type II Supernovae. To facilitate neutrino-induced fission,
the proposed $r$-process scenario is restricted to occur in a
low-density environment such as the neutrino-driven wind from the
neutron star. Further studies to develop this
scenario are emphasized.
\end{abstract}

\section{Introduction}
The nuclei above $^{56}$Fe are produced dominantly by
the slow ($s$) and rapid ($r$) neutron-capture ($n$-capture) processes.
The main $s$-process occurs in low-mass asymptotic giant branch
stars and mainly produces the nuclei above $^{88}$Sr while
the weak $s$-process occurs in massive stars and mainly produces
the nuclei up to $^{88}$Sr (e.g., K\"appeler, Beer, \& Wisshak 1989). 
It is considered here that Type II 
supernovae (SNe II) resulting from evolution of massive stars
are the major sources for the $r$-process (e.g., Qian 2000). 
As massive stars have much shorter lifetimes than low-mass
stars, SNe II made the dominant contributions to the abundances 
of $n$-capture elements in metal-poor (MP) stars that were 
formed in the early Galaxy. 
Detailed abundance patterns covering many $n$-capture
elements have been obtained for a number of ultra-metal-poor
(UMP) stars with [Fe/H]~$\approx -3$ (e.g., Sneden et al. 2000;
Westin et al. 2000; Hill et al. 2001). This Letter discusses
the implications of these abundance patterns for $r$-process 
nucleosynthesis in SNe II.

Starting with a large abundance ratio of neutrons relative to 
the seed nuclei (i.e., a large neutron-to-seed ratio), 
the $r$-process produces a distribution of neutron-rich progenitor 
nuclei far from stability through the interplay of rapid 
neutron capture, photo-disintegration, and $\beta$-decay.
Depending on the neutron-to-seed ratio,
the $r$-process may produce progenitor nuclei that subsequently 
undergo fission on a short timescale.
This limits the range of nuclei involved in the
$r$-process and may result in a cyclic flow between
the terminating nuclei and their fission products
(i.e., fission cycling). The role of 
fission in $r$-process nucleosynthesis was discussed 
by the very first papers that proposed the 
$r$-process (Burbidge et al. 1957; Cameron 1957).
Fission cycling was discussed in detail by Seeger
Fowler, \& Clayton (1965) and was included in many
$r$-process studies (e.g., Rauscher et al. 1994;
Freiburghaus, Rosswog, \& Thielemann 1999; Cameron 2001). Fission 
was also invoked to explain certain features of the observed 
$r$-process abundance patterns ($r$-patterns). For example,
Cameron (1957) attributed the small peak in
the rare earth region (with mass numbers $A\sim 160$) of the solar 
$r$-pattern to fission of the progenitor nuclei with $A\sim 287$.

An $r$-process scenario with fission but no fission cycling is 
considered here to account for the observed abundance patterns
in UMP stars 
(\S2). It is proposed that fission may be induced by neutrino 
reactions with the progenitor nuclei after the $r$-process freezes 
out (i.e., exhausts all the initial neutrons) in SNe II. The 
conditions under which this scenario could be realized and some
possible features of the resulting $r$-pattern
are discussed (\S3). Further studies 
to develop this scenario are emphasized (\S4).

\section{Abundance Patterns in UMP Stars}
Extensive observations
showed that the abundances of the heavy $n$-capture elements
from Ba up to Pt in UMP stars exhibit remarkable regularity and
follow the solar $r$-pattern rather closely (e.g., Sneden et al. 
2000; Westin et al. 2000). However, as shown in Figure 1a, 
the observed abundances of the light $n$-capture 
elements Rh, Pd, Ag, and Cd in the UMP star CS 22892-052 
(Sneden et al. 2000) are too low relative to the solar $r$-pattern 
(Arlandini et al. 1999) that is translated to fit the region above Ba 
($A>130$). A possible explanation for this
is that different production mechanisms may be responsible for the
$n$-capture elements below and above Ba in UMP stars. For 
example, fission cycling tends to produce a rather robust $r$-pattern 
at $A\gtrsim 130$ (e.g., Freiburghaus et al. 1999).
So the regular abundance pattern for Ba and above in UMP stars could 
be attributed to an $r$-process with a neutron-to-seed ratio that is
sufficiently high for fission cycling to occur. Then the light 
$n$-capture elements in UMP stars could be attributed to 
some other $r$-process with a neutron-to-seed ratio that is 
appropriate for producing these elements only or to the weak 
$s$-process that might operate at some unusually high efficiency 
in the earliest generations of massive stars.

However, the attribution of the abundances of Ba and above in UMP 
stars to an $r$-process with fission cycling may be in conflict with
the data on another UMP star CS 31082-001. As can be 
seen from Figure 1b, the observed abundances in CS 31082-001 
(Cayrel et al. 2001; Hill et al. 2001) and CS 22892-052 
(Sneden et al. 2001) follow approximately the same pattern except 
that the differences in the abundances of
Os and the radioactive elements Th and U between the two stars 
appear to be significantly larger than those for the other 
elements (the differences in the abundances of Ir and Pb may also
be significant but the uncertainties are large).
As can be best seen by considering the abundance 
ratio Th/Eu, the data on CS 31082-001 and CS 22892-052 do not
support a robust $r$-pattern at $A\gtrsim 130$ that is 
expected from fission cycling. 
In view of the extremely low [Fe/H] values 
of $\approx -3$ for the two stars, it is reasonable to assume that 
both stars were formed during the first few Gyr after the onset of 
Galactic $r$-process nucleosynthesis. So both stars should have
ages of $\sim 10$ Gyr. Under the assumption of a fixed $r$-process 
yield ratio of Th relative to Eu, the difference $\Delta t$ in the 
ages of the two stars is 
\begin{displaymath}
\Delta t={\tau_{232}\over\log e}\left|
\log\left({{\rm Th}\over {\rm Eu}}\right)_{\rm CS\ 31082}-
\log\left({{\rm Th}\over {\rm Eu}}\right)_{\rm CS\ 22892}\right|,
\end{displaymath}
where $\tau_{232}=20.3$ Gyr is the lifetime of $^{232}$Th. 
The data $\log\epsilon({\rm Eu})=-0.70\pm 0.09$,
$\log\epsilon({\rm Th})=-0.96\pm 0.03$ for CS 31082-001 
(Cayrel et al. 2001; Hill et al. 2001) and 
$\log\epsilon({\rm Eu})=-0.93\pm 0.09$,
$\log\epsilon({\rm Th})=-1.60\pm 0.07$ for CS 22892-052 
(Sneden et al. 2001), where 
$\log\epsilon({\rm E})=\log{\rm (E/H)}+12$,
give $\Delta t=19.2\pm 7.0$ Gyr. This is totally unreasonable. Thus,
the assumption of a fixed $r$-process yield ratio of Th relative to 
Eu must be invalid and the $r$-pattern at $A\gtrsim 130$ is not robust.

While it may be possible to accommodate the data on both CS 22892-052 
and CS 31082-001 by having variable amounts of fission cycling in
the $r$-process, an alternative scenario with fission but no fission 
cycling is considered here to explain the observed abundance patterns 
of $n$-capture elements in these two and possibly other 
UMP stars. It is proposed that an $r$-process produces a freeze-out
distribution of progenitor nuclei covering 
$190\lesssim A<320$ with a peak at $A\sim 195$ (corresponding
to the magic neutron number $N=126$). The required neutron-to-seed ratio
is high, but it is assumed that no fission cycling occurs during the 
$r$-process. This may be
consistent with some recent studies of fission barriers for 
extremely neutron-rich nuclei (Mamdouh et al. 2001). As the progenitor
nuclei $\beta$-decay towards stability after freeze-out, all of those
with $260\lesssim A<320$ eventually undergo spontaneous
fission (e.g., Cameron 2001). Due to the strong influence of the closed
proton and neutron shells at $^{132}$Sn, the fission of
$260\lesssim A<320$ is expected to produce one fragment at
$A\sim 132$ and the other at $130\lesssim A<190$ although the detailed
fission yields may be more complicated (e.g., Hulet et al. 1989). 
Some of the
progenitor nuclei with $230\lesssim A<260$ would also undergo 
spontaneous fission during decay towards stability, thereby producing 
one fragment again at $A\sim 132$ and the other at $100\lesssim A<130$.
The possibility of fission for 
$230\lesssim A<260$ may be greatly enhanced by reactions with the
neutrinos emitted in SNe II as these nuclei could be highly excited
by such reactions. Neutrino reactions may even induce fission of
the progenitor nuclei with $190\lesssim A<230$. Experiments using
energetic particles to induce fission of the stable or long-lived
nuclei in this mass range showed that the fission mode is dominantly
symmetric with no preference for a
fission fragment at $A\sim 132$ and the mass ratio of the two fission 
fragments is $\sim 1$--1.2 (e.g., Britt et al. 1963; see 
M\"oller et al. 2001 for recent theoretical interpretation). 
So the neutrino-induced
fission of $190\lesssim A<230$ is expected to produce fragments at
$86\lesssim A<125$. 

Thus, in the $r$-process scenario proposed here,
the fission of the progenitor nuclei with $190\lesssim A<320$ during
decay towards stability after freeze-out
would produce nuclei with $86\lesssim A<190$ (see Table 1). 
In view of the mass
range of the progenitor nuclei and their fission products, it is
tempting to consider the proposed $r$-process scenario as the
dominant source for all the $n$-capture elements observed in
UMP stars. The ratio of the fission yields at $86\lesssim A<190$ 
relative to the 
surviving abundances at $190\lesssim A<260$ depends on the number 
of neutrino reactions experienced by each progenitor nucleus after 
freeze-out. Variation of this number among individual SNe II
may explain why the observed abundances of Os and above 
($A\gtrsim 190$) relative to those of the $n$-capture elements
below Os are significantly different in
CS 22892-052 and CS 31082-001. The conditions 
under which the proposed scenario could be realized are discussed
below.

\section{Neutrino-Induced Fission and $r$-Process Nucleosynthesis}
In the $r$-process scenario proposed here, all the initial neutrons 
are exhausted when the heaviest progenitor nuclei with $A\sim 320$ 
are produced. Except for the progenitor nuclei perhaps with 
$300\lesssim A<320$ that are mostly produced when the $r$-process 
is running out of neutrons, the freeze-out abundances of the other 
progenitor nuclei are expected to be inversely proportional to their 
$\beta$-decay rates. The $\beta$-decay rates for the progenitor nuclei 
with $190\lesssim A<200$ ($N=126$) are $\sim 10$ times smaller 
than those for 
$200\lesssim A<300$ (M\"oller et al. 1997). With this crude guidance
to the freeze-out abundances (see Table 1), the ratio of the fission 
yields at $86\lesssim A<190$ relative to the surviving abundances at 
$190\lesssim A<260$ is $\sim (2f+0.5):(1-f)$ if a fraction $f$ of
the progenitor nuclei with $190\lesssim A<260$ and all of those with 
$260\lesssim A<320$ undergo fission during decay towards stability 
after freeze-out. By assigning the $n$-capture
elements below Os to the fission products and considering Pb as the
dominant decay product of the surviving nuclei with 
$210\lesssim A<260$, it can be estimated from 
Figure 1 that $f\sim 40\%$ is required
to account for the gross features of the observed abundance pattern
in CS 22892-052 and $f\sim 20\%$ for CS 31082-001.

Without excitation by energetic particles, the probability of fission 
increases with $Z^2/A$, where $Z$ is the atomic number of the nucleus.
All of the progenitor nuclei with $260\lesssim A<320$ are expected
to undergo spontaneous fission eventually as $Z$ increases during 
their decay towards stability (e.g., Cameron 2001). 
In contrast, as the stable or long-lived nuclei with
$190\lesssim A<230$ do not undergo any significant spontaneous 
fission, neither would their neutron-rich progenitors in the
absence of high excitation. The probability of
spontaneous fission during decay towards stability may not be 
significant even for the progenitor nuclei with $230\lesssim A<260$
(e.g., Meyer et al. 1989). However, the progenitor nuclei and their
daughters can be highly excited by reactions with the neutrinos
emitted in SNe II. The typical excitation energy is $\sim 30$ MeV
for $\nu_e$ capture and $\sim 15$ MeV for reactions with
$\nu_\mu$, $\bar\nu_\mu$, $\nu_\tau$, and $\bar\nu_\tau$ 
(e.g., Qian et al. 1997). Thus, neutrino reactions may be crucial 
in inducing the fission of $\sim 20$--40\% of the progenitor nuclei 
with $190\lesssim A<260$ during decay towards stability after
freeze-out. Note that neutrino-induced fission should be unimportant
during the $r$-process as the rates for $\nu_e$ capture must be
severely limited in order to produce the abundance peak at 
$A\sim 195$ (Fuller \& Meyer 1995). An essential question is then 
whether there would be a sufficient number of neutrino reactions 
with the progenitor nuclei after the $r$-process freezes out in 
SNe II.

The gravitational binding energy of the neutron star formed in
SNe II is radiated in neutrinos over a period of $\sim 20$ s
(e.g., Woosley et al. 1994). This is the upper limit on the
duration of the $r$-process so that the progenitor nuclei can
have a significant number of neutrino reactions after freeze-out.
The duration of the $r$-process is dominantly controlled by the
$\beta$-decay lifetimes of the progenitor nuclei. For an 
$r$-process path in a low-density environment where electrons
are nondegenerate, the typical $\beta$-decay lifetimes are
$\sim 0.1$ s for $\sim 10$ progenitor nuclei with closed neutron
shells and $\sim 0.01$ s for those without (M\"oller et al. 1997).
It would then take $<4$ s to reach the progenitor nuclei with 
$A\sim 320$. So all the progenitor nuclei could have a significant
number of neutrino reactions
after freeze-out if the $r$-process
scenario proposed here is restricted to occur in a low-density
environment such as the neutrino-driven wind from the neutron
star (e.g., Woosley et al. 1994) instead of a high-density
environment such as an accretion disk (Cameron 2001).

Another concern is 
the probability of fission following a neutrino reaction. A
neutron-rich nucleus far from stability can de-excite by
emitting neutrons instead of undergoing fission. Increase in
$Z$ would help neutrino-induced fission compete with neutron
emission. Based on the $\beta$-decay lifetimes estimated by 
M\"oller et al. (1997), $Z$ could increase by $\sim 8$ units
within $\sim 10$ s for the progenitor nuclei with 
$190\lesssim A<260$. Thus, the crucial role of neutrino-induced
fission in the $r$-process scenario proposed here relies on
the assumption that the probability of fission following a
neutrino reaction is or becomes significant during the first
$\sim 8$ steps in the $\beta$-decay chains of the progenitor 
nuclei with $190\lesssim A<260$. This assumption should be
checked by detailed calculations in the future.
If a typical probability of 
fission is $P_f\sim 50\%$ (i.e., comparable to that of neutron
emission), a fraction $f\sim P_f[1-\exp(-n)]\sim 20$--40\%
of the progenitor nuclei would undergo fission after a total
number $n\sim 0.5$--1.6 of neutrino reactions per nucleus.
The values of $P_f$ to give $f\sim 20$--40\% may be reduced
somewhat by increasing $n$ to $\sim 3$. A similar level of
neutrino reactions would completely account for the
production of $A=183$--187 by neutrino-induced neutron emission
from the progenitor nuclei in the peak at $A\sim 195$
after freeze-out (Qian et al. 1997; Haxton et al. 1997).

Given that the proposed $r$-process scenario with fission
could be realized, it is interesting to note some features
of the corresponding yield pattern that are not so evident from
the data for CS 22892-052 and CS 31082-001. 
As proposed, the freeze-out distribution of progenitor
nuclei covers $190\lesssim A<320$ with a peak at $A\sim 195$.
The fission of $\sim 20$--40\% of the progenitor nuclei with 
$190\lesssim A<260$ and all of those with $260\lesssim A<320$
would produce enhanced fission yields at $A\sim 90$
and $A\sim 132$. The enhancement at $A\sim 90$
is expected from fission of the progenitor nuclei in 
the peak at $A\sim 195$, while that at $A\sim 132$ from the 
common fission fragment at this position for all the progenitor 
nuclei with $230\lesssim A<320$. Thus, the typical yield
pattern from the proposed $r$-process scenario has a peak at
$A\sim 195$ and enhanced yields at $A\sim 90$ and $A\sim 132$.
This yield pattern could be
modified by capturing the neutrons released from 
fission. For example, the positions of the peak and the enhanced
fission yields could be shifted to somewhat higher $A$. 
Such shifts
in the yield pattern due to residual neutron capture were found
in $r$-process scenarios with fission cycling (Rauscher et al.
1994; Freiburghaus et al. 1999). In any case, the $r$-process
scenario proposed here could not account for the
peak at $A=130$ in the solar $r$-pattern and therefore could
provide one of the diverse sources for the $r$-process 
that are required by the model of Wasserburg, Busso, \& Gallino
(1996) to explain the meteoritic data on $^{129}$I and
$^{182}$Hf. The yield pattern resulting from this
scenario might also be similar to the $r$-pattern that was 
derived by Qian \& Wasserburg (2001) for the high-frequency kind 
of SNe II based on the model of Wasserburg et al. (1996) and 
observations of MP stars.

There may be a hint for the shift of the enhanced fission yields
at $A\sim 132$ to higher $A$ from studies of the Ba isotopic
composition in the UMP star HD 140283.
The three $r$-process isotopes of Ba are $^{135}$Ba, 
$^{137}$Ba, and $^{138}$Ba. The odd-$A$ isotopes of an
element have finite nuclear magnetic moments that lead to
hyperfine structure of the atomic spectra. Magain (1995)
performed a detailed analysis of the Ba spectra for HD 140283 and 
concluded that the fraction of the odd-$A$ Ba isotopes is 
$0.08\pm 0.06$, corresponding to abundance ratios 
$^{138}$Ba/$^{135}$Ba and $^{138}$Ba/$^{137}$Ba of $>6$.
By capturing the neutrons released from fission, the enhanced
fission yields at $A\sim 132$ could be shifted to produce this
steep rise of the yield at $A=138$. It is also possible that
the Ba in HD 140283 may not be of pure $r$-process origin. In
any case, the Ba isotopic composition in UMP stars has important
implications for $n$-capture processes in the early Galaxy
and deserves further observational studies.

\section{Discussion and Conclusions}
Clearly, much further work is required to develop the
$r$-process scenario with fission as proposed here. 
On the observational side, extensive
studies of the abundance patterns of $n$-capture elements should
be carried out for many more UMP stars to identify regularities and 
variations. The paucity of data on the light $n$-capture elements
below Ba should be remedied. As Pb is the dominant decay product of
the progenitor nuclei with $210\lesssim A<260$, accurate determination
of its abundance should be made.
Detailed analyses of Ba isotopes are also valuable as discussed in \S3. 
On the theoretical side, the foremost issue to investigate is
the probability of neutrino-induced fission for extremely neutron-rich
nuclei and the corresponding fission yields. Numerical calculations 
with an extensive network including neutrino reactions are needed to 
make a detailed comparison between the theoretical yields and the 
observed abundance patterns. The effects of the energy release from 
fission on the thermodynamic and hydrodynamic evolution of the 
$r$-process environment are also interesting to study.
Last but certainly not the least, attempts should be made to 
understand how the conditions for the proposed $r$-process scenario 
are realized in a low-density environment such as
the neutrino-driven wind in SNe II.

A crucial quantity characterizing the $r$-process conditions is
the neutron-to-seed ratio $R$. By mass conservation, the
average mass number $\langle A\rangle$ for the freeze-out distribution
of progenitor nuclei is $\langle A\rangle=A_{\rm sd}+R$, 
where $A_{\rm sd}$ is the mass number of the seed nuclei. With 
$A_{\rm sd}\sim 90$ obtained in the neutrino-driven wind (e.g.,
Woolsey et al. 1994), 
$R\sim 130$ is required to produce $\langle A\rangle\sim 220$ 
in the proposed $r$-process scenario (see Table 1). In general, $R$ is 
determined by the electron fraction $Y_e$, the entropy $S$, and the 
dynamic timescale $\tau_{\rm dyn}$ of the wind. 
While it remains unclear how
the suitable combinations of $Y_e$, $S$, and $\tau_{\rm dyn}$ can be
obtained in the wind to give $R\sim 130$, several possibilities
have been suggested. One is to consider the general 
relativistic effects of a more massive or compact neutron star.
As first shown by Qian \& Woosley (1996) and later confirmed by
other studies (Cardall \& Fuller 1997; Otsuki et al. 2000;
Thompson, Burrows, \& Meyer 2001), such effects increase $S$ and
decrease $\tau_{\rm dyn}$, both favoring higher $R$. Another possibility
is to consider the effects of neutrino flavor mixing. The $Y_e$ in
the wind is mostly determined by the competition between
$\bar\nu_e+p\to n+e^+$ and $\nu_e+n\to p+e^-$, and therefore is
extremely sensitive to the differences between the luminosities and 
energy spectra of $\bar\nu_e$ and $\nu_e$ (Qian et al. 1993).
As $\bar\nu_\mu$ and $\bar\nu_\tau$ have much higher average energy
than $\bar\nu_e$ and $\nu_e$, mixing between 
$\bar\nu_{\mu(\tau)}$ and $\bar\nu_e$ could significantly reduce 
$Y_e$ (e.g., Qian \& Fuller 1995), again favoring higher $R$.
It is plausible that by including both the effects of general relativity 
and neutrino flavor mixing, $R\sim 130$ could be obtained in the wind.
These issues should be examined by future studies.

\acknowledgments

Discussions with Roberto Gallino, Peter M\"oller, and Jerry Wasserburg 
are gratefully acknowledged. I thank the referee, Al Cameron, for
many valuable criticisms.
This work was supported in part by the Department of Energy under
grants DE-FG02-87ER40328 and DE-FG02-00ER41149.

\clearpage
\figcaption{(a) Data on $n$-capture elements
in CS 22892-052 (filled circles; Sneden et al. 2000) compared with 
the solar $r$-pattern (solid line) that is translated to fit
the Eu data. The solar $r$-pattern is derived by subtracting
the stellar-model-based $s$-process contributions from the solar
inventory (Arlandini et al. 1999). 
(b) Data for CS 31082-001 (open circles with that for Pb slightly
shifted for clarity; Cayrel et al. 2001;
Hill et al. 2001) compared 
with those for CS 22892-052 (filled circles). The downward arrows
indicate upper limits.}

\clearpage
\begin{deluxetable}{crr}
\tabletypesize{\footnotesize}
%\tabletypesize{\scriptsize}
%\footnotesize
%\scriptsize
\tablecaption{Fission During Decay Towards Stability}
\tablewidth{0pt}
\tablehead{
\colhead{progenitor nuclei}&
\colhead{freeze-out abundances\tablenotemark{a}}&
\colhead{fission products}
}
\startdata
$190\lesssim A<200$&1&$86\lesssim A<109$\\
$200\lesssim A<230$&$\sim 0.3$&$91\lesssim A<125$\\
$230\lesssim A<260$&$\sim 0.3$&$A\sim 132$ and $100\lesssim A<130$\\
$260\lesssim A<320$&$\sim 0.4$&$A\sim 132$ and $130\lesssim A<190$\\
\enddata
\tablenotetext{a}{Normalized so that
the total freeze-out abundance of $190\lesssim A<200$ (in the peak at 
$A\sim 195$) is unity.}
\end{deluxetable}


\begin{references}

\reference{}
Arlandini, C., K\"appeler, F., Wisshak, K., Gallino, R., Lugaro, M., 
Busso, M., \& Straniero, O. 1999, \apj, 525, 886

\reference{}
Britt, H. C., Wegner, H. E., \& Gursky, J. C. 1963, Phys. Rev., 129, 2239

\reference{}
Burbidge, E. M., Burbidge, G. R., Fowler, W. A.,
\& Hoyle, F. H. 1957, Rev. Mod. Phys., 29, 547 

\reference{}
Cameron, A. G. W. 1957, \pasp, 69, 201

\reference{}
Cameron, A. G. W. 2001, \apj, 562, 456

\reference{}
Cardall, C. Y., \& Fuller, G. M. 1997, \apjl, 486, L111

\reference{}
Cayrel, R., et al. 2001, \nat, 409, 691

\reference{}
Freiburghaus, C., Rosswog, S., \& Thielemann, F.-K. 1999, \apjl, 525, L121

\reference{}
Fuller, G. M., \& Meyer, B. S. 1995, \apj, 453, 792

\reference{}
Haxton, W. C., Langanke, K., Qian, Y.-Z., \& Vogel, P. 1997, \prl, 78, 2694

\reference{}
Hill, V., Plez, B., Cayrel, R., \& Beers, T. C. 2001, astro-ph/0104172

\reference{}
Hulet, E. K., et al. 1989, \prc, 40, 770

\reference{}
K\"appeler, F., Beer, H., \& Wisshak, K. 1989, Rep. Prog. Phys., 52, 945

\reference{}
Magain, P. 1995, \aap, 297, 686

\reference{}
Mamdouh, A., Pearson, J. M., Rayet, M., \& Tondeur, F. 2001, 
Nucl. Phys. A, 679, 337

\reference{}
Meyer, B. S., Howard, W. M., Mathews, G. J., Takahashi, K., M\"oller, P.,
\& Leander, G. A. 1989, \prc, 39, 1876

\reference{}
M\"oller, P., Madland, D. G., Sierk, A. J., \& Iwamoto, A. 2001, 
\nat, 409, 785

\reference{}
M\"oller, P., Nix, J. R., \& Kratz, K.-L. 1997, ADNDT, 66, 131

\reference{}
Otsuki, K., Tagoshi, H., Kajino, T., \& Wanajo, S. 2000, \apj, 533, 424

\reference{}
Qian, Y.-Z. 2000, \apjl, 534, L67

\reference{}
Qian, Y.-Z., Fuller, G. M., Mathews, G. J., Mayle, R. W., Wilson, J. R.,
\& Woosley, S. E. 1993, \prl, 71, 1965

\reference{}
Qian, Y.-Z., \& Fuller, G. M. 1995, \prd, 52, 656

\reference{}
Qian, Y.-Z., Haxton, W. C., Langanke, K., \& Vogel, P. 1997, 
\prc, 55, 1532

\reference{}
Qian, Y.-Z., \& Wasserburg, G. J. 2001, \apj, 559, 925

\reference{}
Qian, Y.-Z., \& Woosley, S. E. 1996, \apj, 471, 331

\reference{}
Rauscher, T., Applegate, J. H., Cowan, J. J., Thielemann, F.-K., \&
Wiescher, M. 1994, \apj, 429, 499

\reference{}
Seeger, P. A., Fowler, W. A., \& Clayton, D. D. 1965, \apjs, 11, 121

\reference{}
Sneden, C., Cowan, J. J., Ivans, I. I., Fuller, G. M., Burles, S., 
Beers, T. C., \& Lawler, J. E. 2000, \apjl, 533, L139

\reference{}
Thompson, T. A., Burrows, A., \& Meyer, B. S. 2001, \apj, 562, 887

\reference{}
Wasserburg, G. J., Busso, M., \& Gallino, R. 1996, \apjl, 466, L109

\reference{}
Westin, J., Sneden, C., Gustafsson, B., \& Cowan, J. J. 2000, 
\apj, 530, 783

\reference{}
Woosley, S. E., Wilson, J. R., Mathews, G. J., Hoffman, R. D., \& 
Meyer, B. S. 1994, \apj, 433, 229

\end{references}
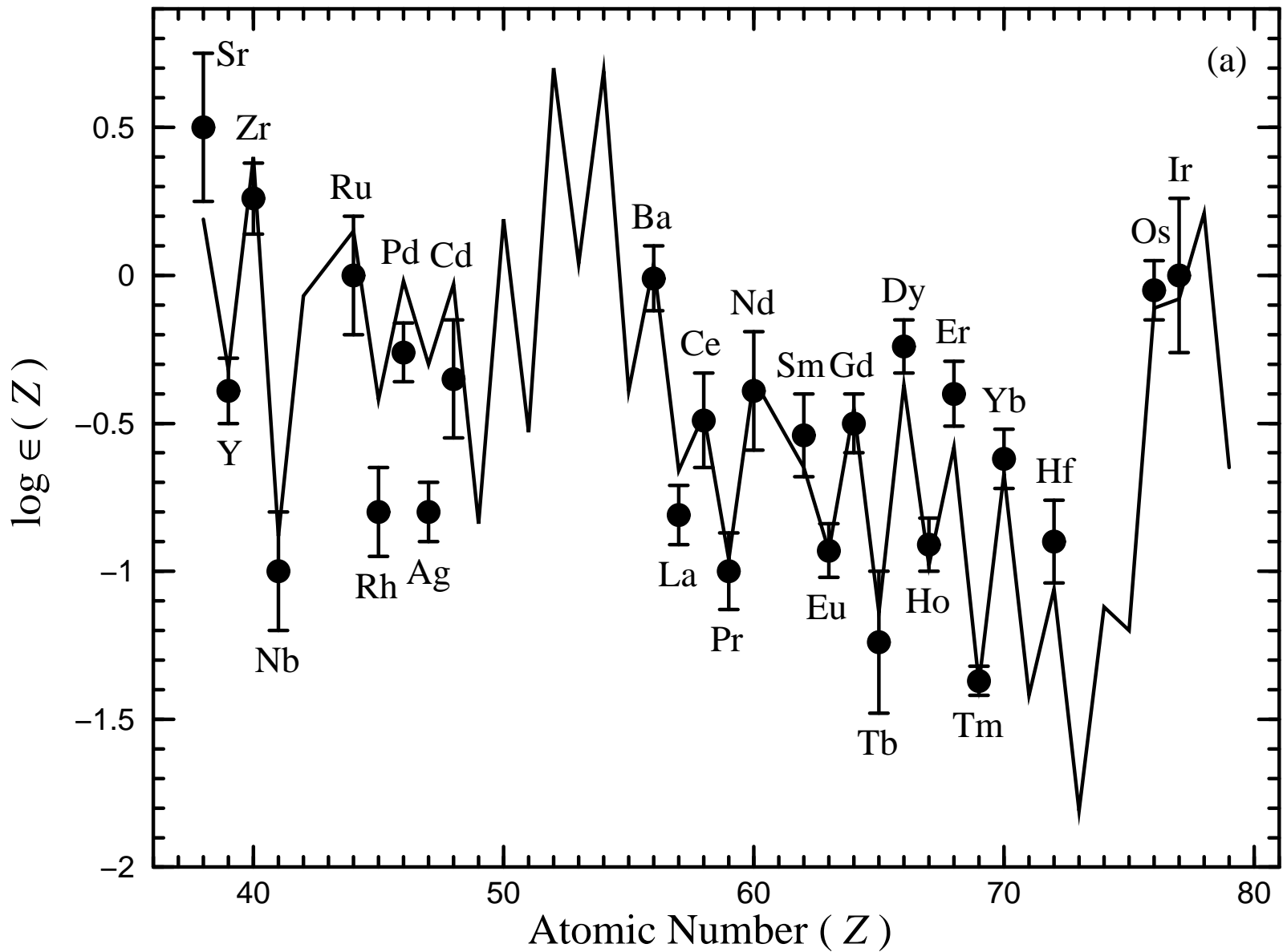
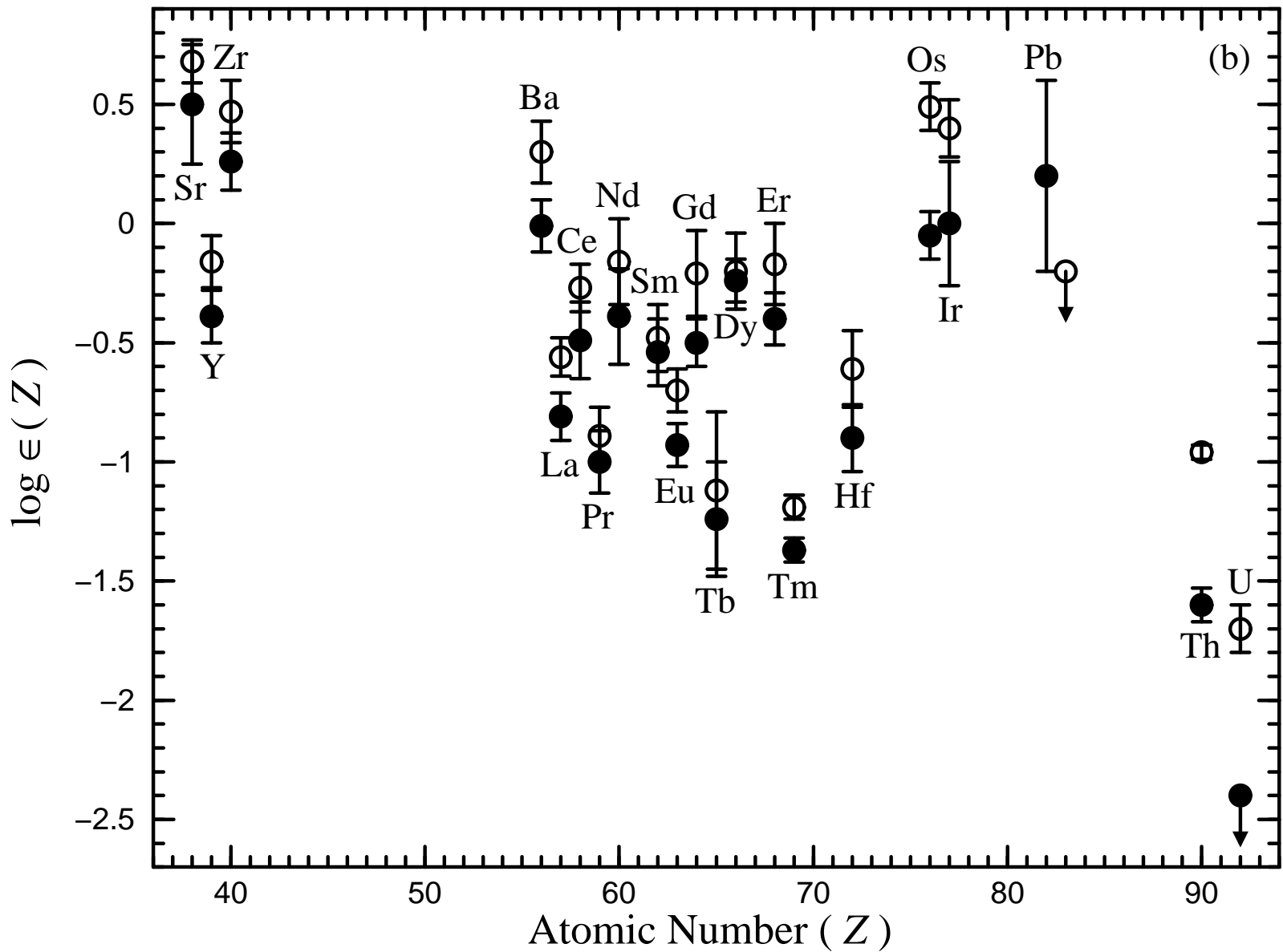
\end{document}